\documentstyle[prl,aps,multicol]{revtex}
\begin{document}
\draft
\title{Comment on ``Theory of metal-insulator transitions in gated 
semiconductors''}
\date{4 January 1999}
\maketitle
\begin{multicols}{2}
In a new paper, Altshuler and Maslov~\cite{altshuler98} suggest a 
model to account for the experimentally observed decrease in resistivity 
with decreasing temperature which has been attributed to a transition to 
an unexpected conducting phase in two-dimensional electron (or hole) 
systems (see, {\it e.g.}, 
Refs.~\cite{kravchenko95,popovic97,coleridge97,hanein98,simmons98,papadakis98,yoon98,hanein98a}).  The mechanism they propose is based on charging and discharging of the 
positively-charged traps which are known to exist in the oxide close to the 
Si-SiO$_2$ interface in silicon MOSFET's.  Within this model, the strong 
temperature dependence and magnetic field dependence of the resistance 
observed in the experiments derive from temperature- and field-induced changes in the charge state of the traps.  The anomalous behavior of these 
materials is thus ascribed to properties of the oxide and the Si-SiO$_2$ 
interface rather than to intrinsic behavior of interacting electrons associated with a 
conductor-insulator transition in two dimensions.

Although the theoretical curves shown in Ref.~\cite{altshuler98} appear 
qualitatively similar to the experimental data~\cite{kravchenko95,popovic97,coleridge97,hanein98,simmons98,papadakis98,yoon98,hanein98a}, close examination reveals significant discrepancies.  For example, the measured resistivity of silicon MOSFET's strongly depends 
on temperature and magnetic field {\em only} at temperatures below 
approximately $\frac{1}{3}T_F$, corresponding to a dimensionless 
temperature $t=0.08$ (at $\epsilon_F/\epsilon_d=0.25$) in Ref.~\cite{altshuler98} (here $T_F$ is the Fermi temperature).  In contrast, the resistivity calculated by Altshuler and Maslov varies with temperature and with magnetic field at temperatures well above $t=0.08$.  In fact, the calculated resistivity is shown in Ref.~\cite{altshuler98} {\em only} for $t>0.05$ ($t>0.1$ in most figures), {\it i.e.}, for the regime where one observes essentially no temperature or magnetic field  dependence of the resistivity.  Moreover, a change in gate voltage of a few percent at $t=0.25$ (corresponding to a ``high'' temperature $T=T_F$) causes more than an order of magnitude change in the calculated resistivity~\cite{altshuler98}, in sharp contrast with experiments~\cite{kravchenko95,popovic97,coleridge97,hanein98,simmons98,papadakis98,yoon98,hanein98a} which show that the resistivity depends so strongly on gate voltage only at low temperatures, $T<<T_F$.

While some of these differences could possibly be remedied by refining the theory, there is a more general and quite fundamental difficulty with the 
Altshuler-Maslov model which we would like to raise in this Comment.  

The model attributes the behavior of a {\em variety of electron and hole systems} to unintended and uncontrolled traps introduced during the 
fabrication of the samples.  One surely expects behavior determined 
by details associated with traps (the number of these traps as well as 
their distribution in energy) to be sample-specific, contrary to what 
is found experimentally.  For example, although samples with different 
spacer material were used in the experiments on p-GaAs/AlGaAs 
heterostructures in Refs.~\cite{hanein98} and~\cite{yoon98}, very 
similar behavior was found for the resistivity.  To illustrate this point 
further, we consider the value of the resistivity, $\rho_c$, at the separatrix between conducting and insulating phases.  In Si MOSFET's~\cite{kravchenko95,popovic97}, p-SiGe heterostructures ~\cite{coleridge97}, p-GaAs/AlGaAs heterostructures~\cite{hanein98,simmons98,yoon98}, n-AlAs heterostructures~\cite{papadakis98}, and n-GaAs/AlGaAs heterostructures~\cite{hanein98a}, this value was found to vary between 0.4 and 
$3\,h/e^2$.  Within the Altshuler-Maslov model, the resistivity at the 
separatrix is determined by two quantities:  the number of traps, variously estimated to be between $10^{10}$~\cite{ando82} and 
$10^{12}\,$cm$^{-2}$~\cite{hori97} in Si~MOSFET's (and unknown in GaAs/AlGaAs heterostructures), and ``critical'' particle density determined by the trap energy level $\epsilon_t$, which is also not known.  It is very unlikely that a strong function (see Eq.~9~a,b in Ref.~\cite{altshuler98}) of two largely unknown parameters which are surely different from sample to sample, and especially from material to material, would yield critical resistivities that differ by less than a factor of ten in the five different systems studied to date.\vspace{3mm}\\
S.~V.~Kravchenko\\
Physics Department, Northeastern University, Boston, Massachusetts 02115\vspace{2mm}\\
M.~P.~Sarachik\\ 
Physics Department, City College of the City University of New York, New York, New York 10031\vspace{2mm}\\
D.~Simonian\\
Department of Physics, Columbia University, New York, New York 
10027\vspace{-2mm}\\

\end{multicols}
\end{document}